# *SwarmFoam*: An OpenFOAM Multi-Agent System Based on Multiple Types of Large Language Models

Chunwei Yang[a], Yankai Wang[a], Jianxiang Tang[a], Haojie Qu[a], Ziqiang Zou[b], YuLiu[b], Chunrui Deng[b], Zhifang Qiu[b, *], Ming Ding[a, *]

[a]Heilongjiang Provincial Key Laboratory of Nuclear Power System & Equipment, Harbin Engineering University, Harbin, 150001 China

[b]State Key Laboratory of Advanced Nuclear Energy Technology, Nuclear Power Institute of China, Chengdu, 610041, China

**Abstract**

Numerical simulation is one of the mainstream methods in scientific research, typically performed by professional engineers. With the advancement of multi-agent technology, using collaborating agents to replicate human behavior shows immense potential for intelligent Computational Fluid Dynamics (CFD) simulations. Some muti-agent systems based on Large Language Models have been proposed. However, they exhibit significant limitations when dealing with complex geometries. This paper introduces a new multi-agent simulation framework, *SwarmFoam*. *SwarmFoam* integrates functionalities such as Multi-modal perception, Intelligent error correction, and Retrieval-Augmented Generation, aiming to achieve more complex simulations through dual parsing of images and high-level instructions. Experimental results demonstrate that *SwarmFoam* has good adaptability to simulation inputs from different modalities. The overall pass rate for 25 test cases was 84%, with natural language and multi-modal input cases achieving pass rates of 80% and 86.7%, respectively. The work presented by *SwarmFoam* will further promote the development of intelligent agent methods for CFD.

## 1. Introduction

CFD is a discipline that uses numerical analysis methods to solve fluid dynamics equations via computers, thus enabling the simulation of physical phenomena. Early research in CFD primarily focused on establishing fundamental theories, such as the Finite Element Method, the Finite Difference Method, and the Lattice Boltzmann Method. With the improvement of computer performance, CFD has been widely applied in fields such as aerospace engineering (Ferro et al., 2023; Lusher et al., 2025), energy and power engineering (Jansasithorn et al., 2025; Xin et al., 2025), and meteorological forecasting (Suárez-Vázquez et al., 2025; Xie et al., 2025), leading to significant advancements in related numerical methods. As Artificial Intelligence (AI) technology matures, the interdisciplinary integration of AI and numerical computation has become one of the latest research topics in the field of CFD.

Prior to 2024, research on AI in CFD primarily concentrated on solving partial differential equation systems, including data-driven methods (Lu et al., 2021; Mao et al., 2023) and physics-informed methods (Raissi et al., 2019). Although some studies utilized deep learning methods for automated agents, these efforts were methodologically limited to specific tasks such as turbulence modeling (Novati et al., 2021) and mesh generation (Kim et al., 2025). In application, they were restricted to single scenarios (Gbadago et al., 2025) and could not be applied to a wide range of CFD simulations.

With the increasing maturity of generative AI technology, Large Language Models (LLMs) have opened up a new research paradigm for CFD. A multi-agent system involves coordinating different agent roles within the same working environment, intelligently invoking external tools according to a predefined workflow to complete specific tasks. This approach has achieved considerable success in complex scenarios, such as autonomous driving (M. Zhang et al., 2025; P. Zhang et al., 2025), clinical diagnostics (Liu et al., 2026; Qiu et al., 2024), and architectural design (Chen & Bao, 2025; Y. Dong et al., 2025). Some scholars have begun to explore the use of LLMs or multi-agent systems to create intelligent agents for the CFD software OpenFOAM, aiming to achieve fluid mechanics simulation relying solely on natural language or high-level instructions.

OpenFOAM is one of the most powerful and widely used open-source CFD platforms. Unlike commercial CFD software, OpenFOAM relies on configuration files to drive all stages of the fluid dynamics simulation and does not depend on a user interactive interface, making it an ideal target for agent-based automation. However, fully automated intelligent CFD workflows face unique challenges due to the complexities of geometry construction and mesh generation, as well as strict constraints on file formats. Consequently, related technical work is currently in a highly competitive and preliminary research and development phase.

In 2024, Tsinghua University released the world's first open-source multi-agent system for CFD, MetaOpenFOAM-1 (Chen et al., 2024). It adopts the MetaGPT (Hong et al., 2023) multi-agent framework, which supports the customization of agent roles and actions. MetaOpenFOAM-1 defined multiple agent roles, acting as proxies for four steps in numerical computation: case analysis, pre-processing, execution of calculation, and error correction. Each agent is equipped with a Q&A interface built via the LangChain programming framework (Jeong et al., 2024) to interact with a local knowledge base containing the OpenFOAM help documentation. In 2025, MetaOpenFOAM-2 was further proposed (Chen et al., 2025). The new version added functionalities for task decomposition (breaking down user requirements into sub-tasks such as simulation, post-processing, and simulation optimization) and post-processing.

In the same year, the University of Exeter in the UK proposed another intelligent simulation system—OpenFOAMGPT(Pandey et al., 2025). OpenFOAMGPT utilized the native capabilities of LLMs to achieve intelligent OpenFOAM workflows without employing a multi-agent framework. Subsequently, OpenFOAMGPT-2 (Feng et al., 2025) was developed, which similarly introduced multi-agent technology by designing specialized agents responsible for pre-processing, prompt generation, software proxy, and post-processing. Compared to the 1.0 version, OpenFOAMGPT-2 significantly improved the success rate of simulation tasks.

Drawing upon the development experience from MetaOpenFOAM and OpenFOAMGPT, Rensselaer Polytechnic Institute proposed the third mainstream FOAM muti-agent system, *Foam-Agent* (Yue et al., 2025a; Yue et al., 2025b). Its workflow is similar to that of the MetaOpenFOAM series, as both utilize LangChain to implement their local knowledge base Q&A interfaces. The distinction lies in the CFD workflow design: *FOAM-Agent* employs another popular multi-agent framework, LangGraph (Wang & Duan, 2024). Compared to MetaGPT, LangGraph provides greater flexibility in agent development but entails higher development complexity. The core of Foam-agent lies in proposing a more scientific method for perceiving and writing multi-file dependencies, where the model references relevant configuration content when generating the current file. This file generation method aligns better with the format characteristics of OpenFOAM, thereby achieving a higher calculation success rate.

However, existing CFD agent systems rely entirely on natural language to drive CFD simulations. When the geometry and physical conditions of the simulation object become more complex, it is difficult to accurately describe the simulation problem relying solely on natural language input. To solve this problem, *SwarmFoam* introduces the Observer agent for the first time. Based on the multi-modal perception mechanism we designed, the Observer agent can understand the simulation task from both image and text modalities of input, which more closely resembles a real simulation process. Furthermore, *SwarmFoam* includes other agents responsible for various stages of the proxy CFD simulation, including pre-processing, running, error correction, and post-processing. The quality of the simulation configuration files is enhanced by a retrieval-augmented generation (RAG) system, from which the agents can obtain information such as reference cases, solver usage help, and file naming conventions. The main contributions of this paper are summarized as follows:

- **Multimodal Perception Mechanism:** In practical work, simulation information is often conveyed in both image and text formats. To effectively process simulation images, we propose a multimodal perception mechanism tailored for CFD simulations and introduce a new Observer agent. This agent combines natural language input to parse the geometric and physical information within the image, which is used to support the creation of mesh files.

- **First-Error-Priority Intelligent Correction Mechanism:** In the process of simulation error correction, we adopt a first-error-priority strategy. This allows the multiple agents to analyze and correct the file most likely to have caused the current simulation round to fail, thus avoiding the analysis of secondary errors derived from a single root cause. This strategy demonstrates a significant cost advantage when handling complex simulation tasks, as it can effectively reduce token usage.

- **Ablation Study and Error Analysis:** We conducted an ablation study on key mechanisms and performed a comparative analysis of the two proposed multimodal perception mechanisms, highlighting the critical role of pre-parsing images in enhancing overall simulation performance. We also discussed the simulation performance when the intelligent error correction mechanism was disabled. We performed a statistical analysis of all simulation errors encountered during the iteration process, analyzed the main error causes, and discussed some of the limitations of LLMs when handling CFD simulation tasks.

The main structure of this paper is as follows. In Section 2.1, we outline the intelligent simulation workflow architecture of *SwarmFoam*, demonstrating the division of roles among the agents. In Section 2.2, the details of the RAG system are introduced. In Section 2.3, the roles and action design, information interaction mechanisms of the various agents within *SwarmFoam* are detailed. In Section 2.4, a complete demonstration of the intelligent simulation under different modalities of input is provided. Section 3 presents the experiments we used to evaluate *SwarmFoam*'s performance, including the dataset setup and results discussion. In Section 4, we summarize the entire research content, and in Section 5, we offer suggestions for the future development of multi-agent systems for CFD.

## 2. Methodology

*2.1 Intelligent Simulation Workflow Architecture*

The framework of *SwarmFoam* is illustrated in Figure 1. To effectively automate CFD tasks, *SwarmFoam* is designed with six types of agent roles: Observer, Architect, InputWriter, Runner, Reviewer, and ParaMaster. Specifically,

**(1) Observer Agent:** Parses natural language and case image inputs.

**(2) Architect Agent:** Plans the file structure required to complete the current simulation task based on the information parsed by the Observer Agent.

**(3) InputWriter Agent:** Generates the initial OpenFOAM configuration files or corrects erroneous configuration files.

**(4) Runner Agent:** Executes simulation commands and captures potential error messages.

**(5) Reviewer Agent:** Analyzes error messages and identifies the paths of the incorrect configuration files.

**(6) ParaMaster Agent:** Writes code to invoke the Paraview post-processor to output images.

After receiving the user's simulation requirements, the agents within the aforementioned multi-agent system will autonomously execute the OpenFOAM simulation based on their designed roles, actions, and workflow. These simulation agents collaborate by interacting with information in the environment to complete the simulation task and output the simulation results.

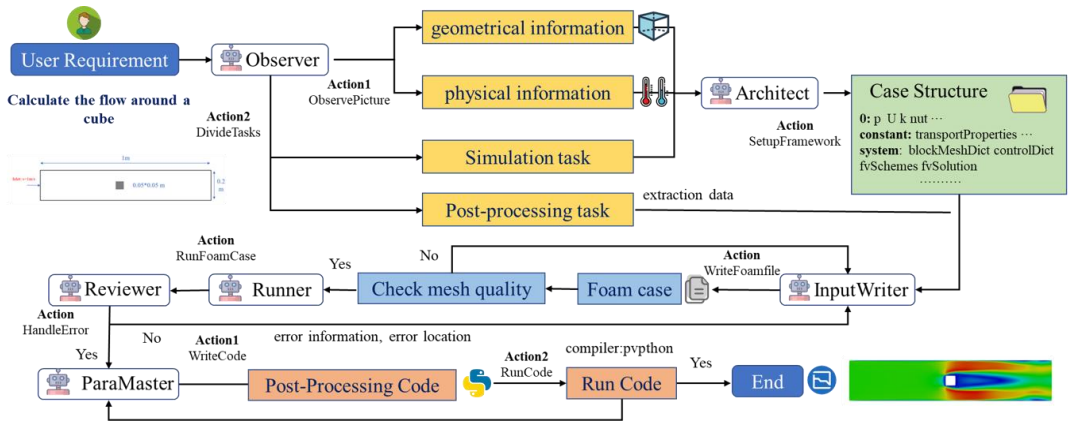

**Fig. 1.** The intelligent simulation workflow of *SwarmFoam*.

*2.2 Retrieval-Augmented Generation System*

*SwarmFoam* supports mainstream text-processing LLMs, such as Deepseek-R1, Deepseek-V3, gemini-2.5-pro, gpt-5-pro, and gpt-4o, as well as multi-modal large language models (MM-LLMs), including gemini-2.5-flash, the Qwen series, and the Llama-4 series. To efficiently handle different categories of tasks, *SwarmFoam* is configured with two types of LLM Q&A interfaces. The first interface type invokes text LLMs for responses based on natural language input. The second interface type invokes multimodal LLMs for Q&A based on multimodal inputs.

As shown in Table 1, *SwarmFoam* utilizes a total of 6 formatted help documents. The first four categories of documents are reference case files: "openfoam_allrun_files" contains 214

OpenFOAM-9 case execution files; "openfoam_cases_struct" contains 227 case structures; "openfoam_commands" includes 84 simulation commands supported by OpenFOAM-9, such as the mesh creation command (blockMesh) and the mesh quality checking command (checkMesh); and "Openfoam_input_files" comprises 3727 case configuration files. The remaining documents contain solver-related content. In OpenFOAM, solvers are used to solve the governing equations and are critical for simulation execution. "solver_describe" provides brief descriptions of the various solvers in OpenFOAM-9, while "solver_help" offers recommendations for the use of each solver, such as the dimension of physical quantities, case structures, appropriate physical problems, and file naming conventions.

The purpose of RAG is to retrieve the aforementioned local expert knowledge and provide it, along with the context, to the LLMs, thereby ensuring the simulation configuration files are more reasonable and standardized (Luo et al., 2026). The basic principle of the RAG technique is illustrated in Figure 2. The entire RAG process is divided into two main steps: indexing and querying. In the indexing stage, the simulation help documents are split into multiple text chunks. These chunks are processed through a loader, a vector model, and a data transformer, and subsequently stored in vector form within multiple local containers. During the querying stage, the vector database retrieves the text segments most similar to the query input. These retrieved segments are then embedded into the prompt for either the Type 1 or Type 2 Q&A interface, depending on the specific task requirements.

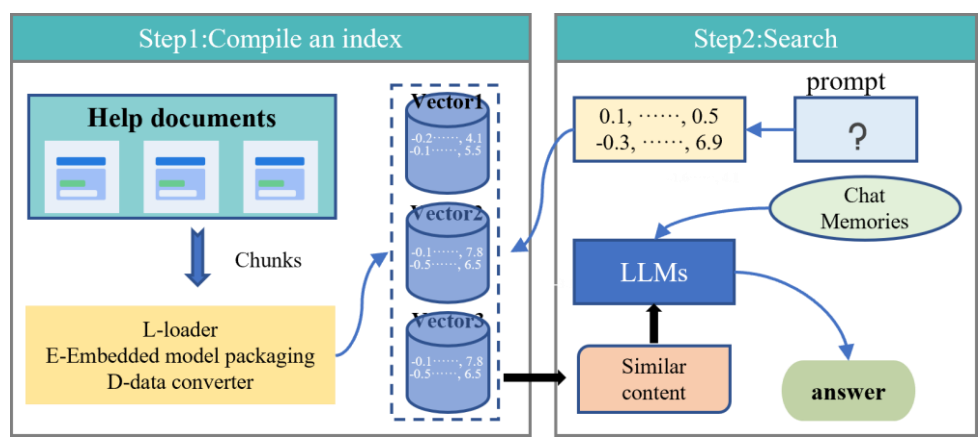

**Fig. 2.** Basic principle of the RAG.

**Table 1**
List of local help documents.

| File name | Content |
| --- | --- |
| openfoam_allrun_flies.txt | Reference for Execution File |
| openfoam_cases_struct.txt | Reference for the composition of case folder |
| openfoam_commands.txt | The simulation commands supported by OpenFOAM-9 |
| openfoam_input_files.txt | Local case configuration file |
| solver_describe.yaml | Introduction to the various solvers of OpenFOAM-9 |
| solver_help.txt | User Guide for OpenFOAM-9 Solvers |

## 2.3 Agent Design

### 2.3.1 Observer Agent

The Observer Agent subscribes only to "UserRequirement" type messages from the environment and publishes the parsed simulation information. When the Observer Agent receives the information published by the user, the agent is triggered. The Observer Agent performs two actions sequentially: ObservePicture and DivideTask. The ObservePicture action is used to parse physical information (e.g., boundary conditions, working fluid type, initial conditions) and geometric information (e.g., geometric shape descriptions, specific dimensions, vertex locations) from the case images. The DivideTask action decomposes the user's requirements into two independent subtasks: simulation and post-processing. If the user input lacks specific post-processing requirements, the Observer Agent will not publish a post-processing task.

The ObservePicture action relies on a simulation-oriented multi-modal perception mechanism. *SwarmFOAM* initially designed two approaches for processing multi-modal information. In the first approach, the image is first deeply parsed by the Observer to capture the embedded simulation information, and is then sent to the InputWriter for configuration file generation. Image parsing and geometry construction are decoupled in this process; the image information is first converted into textual information. In the second approach, the image is directly submitted to the InputWriter along with the prompt for creating the mesh file "blockMeshDict", bypassing the image-to-text conversion process.

The workflow of the first multi-modal perception method is illustrated in Figure 3. The simulation requirements are provided as inputs in two modalities: text and images. Using the bubble rise case as an example, the natural language input is: "Calculate a problem involving the ascent of a 2-dimensional bubble. The total duration is set at 2 seconds." The natural language input does not include specific geometric and physical conditions, as this information is embodied in the image. For the input prompt, a tokenizer and a vector encoder convert it into an embedding vector of a specific dimension. The input image is converted by an image encoder into an embedding vector of the same dimension. Since the vectors share the same dimension, they can be directly concatenated and passed to the MM-LLMs to achieve the parsing of both the image and the natural language.

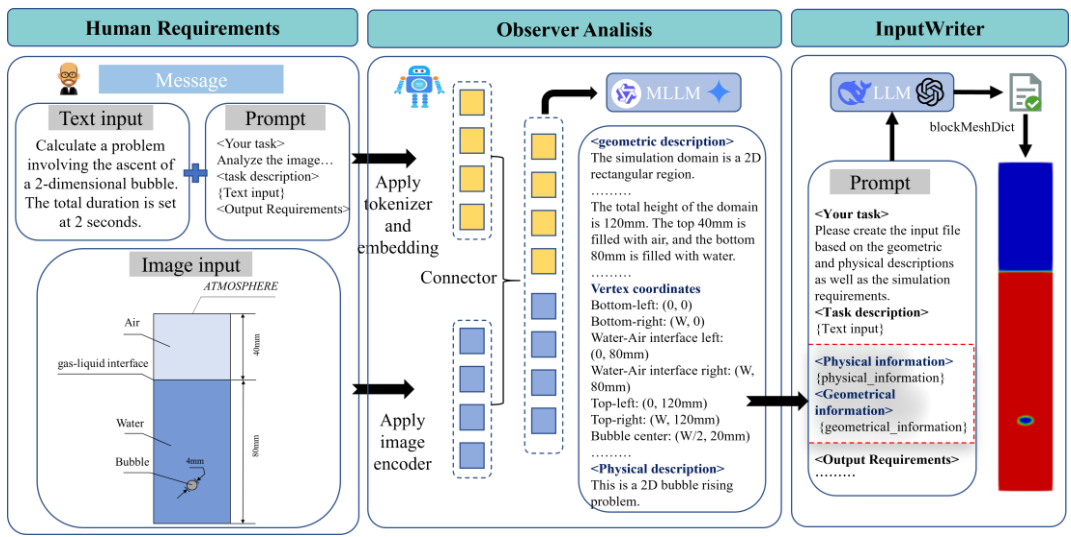

**Fig. 3.** Flowchart for pre-parsing multi-modal information.

After parsing is complete, the Observer accurately describes the current simulation task, including details such as the problem dimensionality, working fluid, geometric dimensions, vertex coordinates, and physical conditions. This information is then embedded into the InputWriter's prompt for the generation of the "blockMeshDict" file.

The workflow of the second multi-modal perception method is illustrated in Figure 4. In this approach, the Observer is responsible only for parsing natural language information, while the image is directly provided to the InputWriter. The multi-modal large language model then simultaneously performs image understanding, simulation requirement comprehension, and mesh file ("blockMeshDict") generation, without any pre-parsing. The advantage of this method is that it simplifies the image processing workflow, but it may lead to the loss of more geometric and physical information.

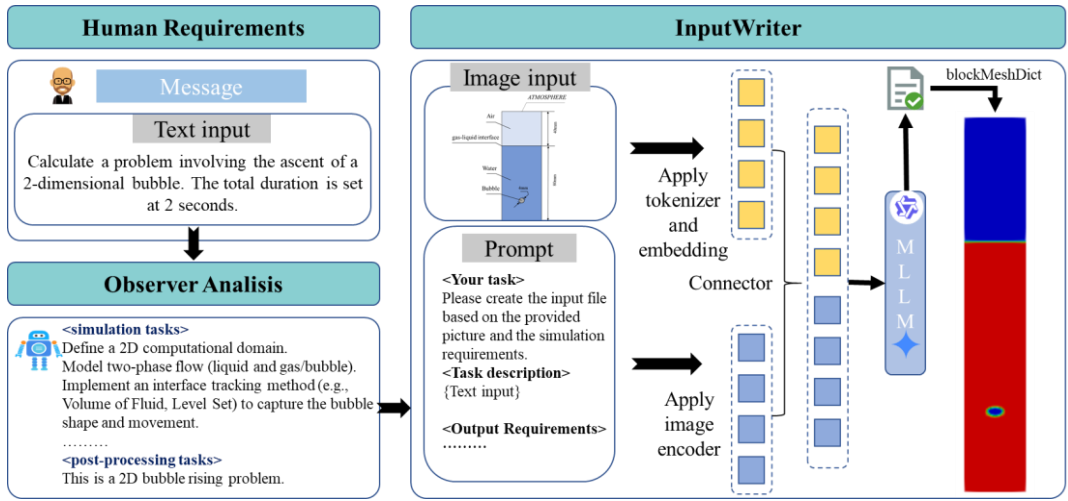

**Fig. 4.** Flowchart for Direct Utilization of Multi-modal Information

In the ablation analysis presented in Section 3.2.3, the impact of pre-parsing on overall simulation performance is discussed. Due to the significant performance disparity, *SwarmFoam* ultimately adopted the first multi-modal perception method.

The multi-modal parsing system can also be utilized in other scenarios requiring image parsing, such as interpreting output images to assess simulation accuracy.

*2.3.2 Architect Agent*

The Architect Agent subscribes to messages published by the Observer Agent and publishes a series of file generation instructions. Upon receiving a message from the Observer Agent, the Architect Agent executes the SetupFramework action. First, the agent parses the current simulation requirement into a structured piece of information containing five elements: case name (e.g., Cavity), case domain (e.g., incompressible), case solver (the solver appropriate for the current problem, e.g., pisoFoam), case category (problem type, e.g., RAS), and solver description (a brief introduction to the solver).

After obtaining this structured information, the Architect Agent retrieves the most similar case structure from the local knowledge base. By combining the retrieval results with the simulation requirements, the Architect Agent defines the OpenFOAM case file structure necessary to complete the current task. A basic case file structure is shown in Table 2. In practice, many other configuration files are often required, depending on the specific problem being solved.

**Table 2**

Basic OpenFOAM Case File Structure

| Folder | File name | Function |
| --- | --- | --- |
| 0 | p | Define the initial and boundary conditions of the pressure field |
| | U | Define the initial and boundary conditions of the flow velocity field |
| constant | transportProperties | Define transport parameters (viscosity) |
| system | blockMeshDict | Define the parameters for grid drawing |
| | controlDict | Define the control parameters of the example calculation |
| | fvSchemes | Definition of the discrete format of parameters |
| | fvSolution | The solution method for defining parameters |
| root | Allrun | Define the execution of the case instruction |

Given the interdependencies among OpenFOAM configuration files, the Architect Agent publishes the file generation instructions sequentially. For files within different directories, the subtask priority is ranked from highest to lowest as follows: "system", "constant", "0", and "Allrun". Within the "system" directory, the subtask priority is: "blockMeshDict", "controlDict", "fvSchemes", "fvSolution", followed by other files. Within the "constant" directory, the priority is "transportProperties", followed by other files. All files within the "0" directory are treated as having the same priority.

*2.3.3 InputWriter Agent*

The InputWriter Agent subscribes to messages published by the Architect Agent and the Reviewer Agent (from the HandleError action), and in turn, publishes simulation execution commands to the environment. The InputWriter Agent's action, WriteFoamfile, is designed as two subtasks: FirstWrite and CorrectFile. Depending on the source of the received message, the InputWriter Agent will select which subtask type to execute.

If the current message is published by the Architect Agent, indicating that the simulation workflow is in the initial configuration file generation stage, the InputWriter Agent executes the FirstWrite subtask. FirstWrite is responsible for generating the series of configuration files listed in Table 2. The generation of each file involves four steps: retrieval from the local knowledge base, checking dependent files, configuration file generation, and file cleaning. During the local knowledge base retrieval step, the agent searches for similar cases and solver help documentation. Checking dependent files aims to ensure consistency in boundary conditions, boundary naming, and unit settings across different configuration files. The file cleaning step is intended to remove residual identifiers (e.g., bash, Foam at the beginning of the file) from the output of LLMs. These identifiers, which do not conform to OpenFOAM file specifications, are a significant cause of simulation failure.

If the current message is published by the Reviewer Agent, the InputWriter Agent executes the CorrectFile subtask. Based on the received error file path and error description, this subtask corrects the erroneous file or creates the missing file. This error correction process also follows the same four steps as FirstWrite.

*2.3.4 Runner Agent*

The Runner Agent subscribes to messages published by the InputWriter Agent and, in turn, publishes either a success flag or error content to the environment. It is a single-action agent that executes the RunFoamCase action. When triggered, the Runner Agent executes the "Allrun" file to start the case simulation, recording the execution sequence of the simulation commands. After the run is completed, a series of log files are generated in the case directory. If the simulation fails, the Runner Agent sequentially captures the error messages from each simulation log file.

*2.3.5 Reviewer Agent*

The Reviewer Agent subscribes to messages from the Runner Agent and is responsible for publishing parsed error messages and their corresponding file paths. This agent possesses two actions: HandleError and EndMark (Algorithm 1). If the simulation completes successfully and post-processing is required, the Reviewer Agent executes the EndMark action, publishing a message directly to the ParaMaster Agent. If the simulation is successful and no post-processing is required, the simulation workflow terminates.

```
Algorithm 1 Error Diagnosis and Correction
Require: Message history H, Configuration C
Ensure: Error diagnosis D, Error file path F
 1: (checkMeshfailed) ← Parse_Input(H, C)                    ▷ Initialize paths and data
 2: LLM_Agent ← Initialize_Agent()                           ▷ Setup LLM interface
 3: if Simulation.successful then                            ▷ Check for no-error signal
 4:     return Next Step, No Error
 5: end if
                                              ▷ — Tier 1: Handle Specific, Known Errors —
 6: if NOT Exists(task_path + /log.blockMesh) then
 7:     (D, F) ← Handle_BlockMesh_Error(task_path)
 8:     return D, F
 9: end if
10: if Exists(Timeprecisionerror) then
11:     (D, F) ← Handle_TimeFormat_Error(task_path)
12:     return D, F
13: end if
                                              ▷ — Tier 2: General Runtime Error Diagnosis —
14: Restore_State(files, task_path)                          ▷ Reset to pre-run state
15: first_err ← errors[0]
16: (D_parsed, F_parsed) ← Parse_Error_Log(first_err)        ▷ Initial error parsing
17: error_type ← Classify_Error_Type(LLM_Agent, first_err, files)  ▷ Use LLM
    to classify
18: if error_type = format error then
19:     (D, F) ← Diagnose_Format_Error(LLM_Agent, ...)
20: else if error_type = Missing file then
21:     (D, F) ← Diagnose_Missing_File(LLM_Agent, ...)
22: else
23:     (D, F) ← (Unknown type error)
24: end if
25: return D, F                                              ▷ Return final diagnosis
```

If the simulation fails, the agent executes the HandleError action. The agent first verifies whether the geometry was successfully created and if the mesh quality is acceptable. If these requirements are not met, it publishes an instruction to regenerate the blockMeshDict file. During the OpenFOAM simulation process, commands such as geometry creation and solver execution are performed sequentially, with each command generating a log file. If the geometry was successfully created, the Inspector Agent deletes the files generated during the run and then proceeds to analyze the content of the received error log.

In contrast to *Foam-Agent* and MetaOpenFOAM, *SwarmFoam* adopts an entirely different error correction principle. In existing systems, the Reviewer Agent aggregates all errors present in the log files, submits them en masse to the large language model for analysis, and subsequently sends a series of modification instructions to the InputWriter Agent. *SwarmFoam*, however, employs a "first-error-priority" principle.

During each correction iteration, the Agent assumes that the first error encountered is the primary cause of the current simulation failure, providing the reason and path only for that specific error. This principle avoids unnecessary analysis of derivative errors triggered by the initial fault. In this analysis process, simulation errors are classified into two major categories: missing files and formatting errors. After completing the analysis, the Reviewer Agent publishes a message to the InputWriter Agent.

Since the messages indicating simulation success and simulation failure originate from different action types, only one specific downstream agent is triggered after the Reviewer Agent completes its task.

*2.3.6 ParaMaster Agent*

The ParaMaster Agent subscribes to messages published by the Reviewer Agent (originating from the EndMark action). *SwarmFoam* implements post-processing functionality by writing Python code to invoke the ParaView program. ParaView is a widely used program for visualizing OpenFOAM cases and provides a callable Python interface, pvpython. Therefore, the ParaMaster Agent in *SwarmFoam* essentially functions as a code generator specifically for ParaView.

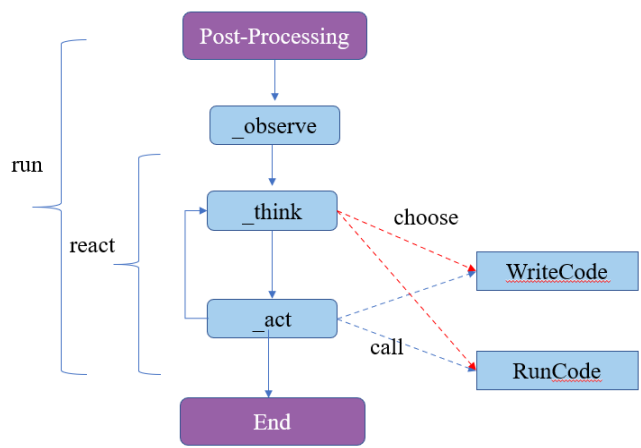

**Fig. 5.** ParaMaster Agent Workflow

As shown in Figure 5, it possesses two actions: WriteCode and RunCode. WriteCode is responsible for writing and correcting code, while RunCode is used for code execution. The Post-processing Agent employs a reasoning mechanism to determine which action type to execute in each iteration. If, during this reasoning process, the agent determines that no further actions are necessary, the entire post-processing workflow concludes, and the resulting images are outputted. Once post-processing is finished, the entire agent-based simulation process terminates, as there are no further messages in the environment to be subscribed to.

## 2.4 Simulation Demonstration

### 2.4.1 Natural Language Input

Figure 6 demonstrates the agent-based simulation for a multi-component combustion case (methane jet), where the simulation requirements are provided via natural language input. The user's natural language input provides a brief description of the computational conditions (e.g., basic geometry, injection port, flow direction, and working fluid type). Since the input does not include images, the Observer Agent bypasses the image parsing process; consequently, the "Geometric description" and "Physical description" fields are both set to 'None'.

Based on the user input, the Observer Agent conducts a detailed analysis of the current simulation requirements (designated as "simulation tasks"), which includes solver selection, initial conditions, and the application of boundary conditions. The Observer Agent then sends the parsed simulation information to the Architect Agent. The Architect Agent performs preliminary planning to determine the necessary configuration files for simulating the multi-component combustion problem. These include files for the initial mass fraction fields (CH4, N2, O2); the initial pressure ($p$), temperature ($T$), and velocity ($U$) fields; as well as files for physical properties, chemical reactions, thermophysical properties, the mesh, and solver settings.

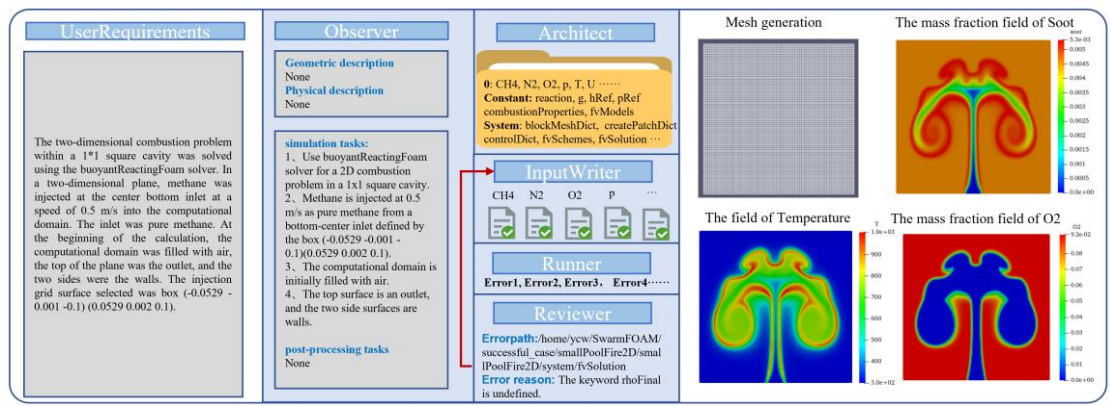

**Fig. 6.** Simulation Demonstration for Natural Language Input

Once the case architecture is established, the InputWriter agent sequentially generates the configuration files and places them in their appropriate paths. The Runner agent executes the "Allrun" file to run the simulation case and sequentially captures a series of errors from the terminal and log files. The Reviewer agent analyzes the first error. As shown in the figure, this agent determines that the critical reason for this simulation failure is the missing "rhoFinal" parameter in the pressure equation settings, and it provides the path to the file causing this error (fvSolution). This error information is sent to the environment. In the next iteration, the InputWriter agent will rewrite the "rhoFinal" keyword setting in the "fvSolution" file.

Through iterative error correction, *SwarmFoam* successfully simulated the multi-component combustion of a jet in a two-dimensional chamber. At $t = 0.9$s, the results show that the soot mass fraction, temperature, and oxygen concentration exhibit similar spatial distributions. As methane is continuously injected, the combustion process gradually intensifies, causing the flame to propagate and generate flame vortices. In the combustion zone, the oxygen concentration decreases, the temperature rises, and soot is produced.

*2.4.2 Multimodal Input*

Figure 7 illustrates the simulation demonstration for a case where the requirements were provided using multimodal inputs (both natural language and images). The figure depicts a case of airflow over an obstacle. After parsing both the image and the natural language, the Observation Agent provided the geometric and physical information required for the simulation.

In the geometric information, the Observation Agent identified the computational domain as a two-dimensional rectangle with a length of 1200 mm and a height of 200 mm. An obstacle was placed on the bottom boundary of the computational domain, and its geometric dimensions and position were accurately described. After this initial parsing of the overall geometry, the Observer further examined each boundary in detail, providing their coordinate ranges and types. As shown in the figure, the Observer identified the left boundary of the domain as 'Inlet', the right boundary as 'Outlet', and the remaining boundaries as 'Wall'. These observed boundary conditions are consistent with the input case image.

In the physical information, the Observer identified the problem as a 2D, isothermal, turbulent, incompressible flow case to be simulated using the pisoFoam solver. It further provided details such as the working fluid type, flow velocity, temperature, and pressure.

The intelligent agent workflow following the observation phase is similar to that shown in Figure 6. *SwarmFoam* planned the required file types and generated the configuration files. After execution, the agent reported the cause of the error and its corresponding path. Through iterative error correction, *SwarmFoam* successfully simulated the obstacle flow process. As per the post-processing requirements, the ParaMaster Agent generated the velocity and pressure field distribution plots. From these results, it can be seen that due to the reduction in the flow cross-section, a significant acceleration effect occurred in the upper region downstream of the obstacle, while a low-velocity zone formed in the lower region.

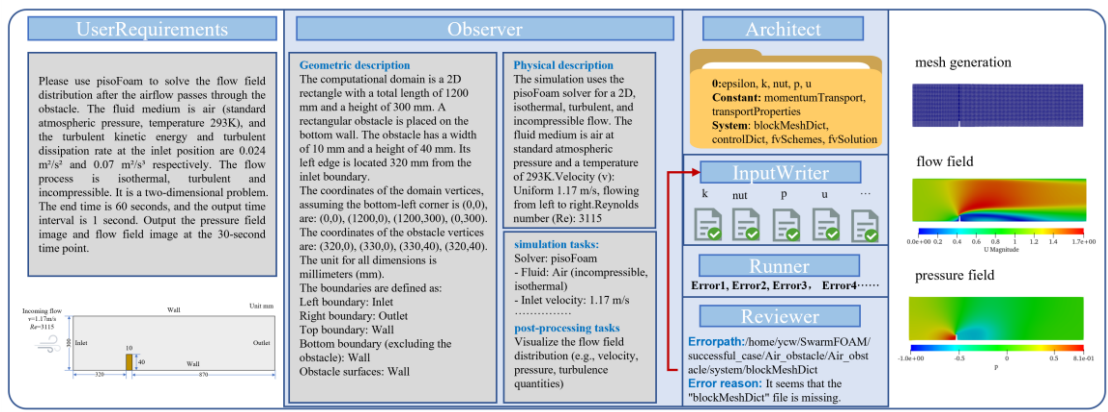

**Fig. 7.** Simulation Demonstration for Multimodal Input

# 3. Experiments test and Application

## 3.1 Experimental Setup

### 3.1.1 Dataset

As shown in Table 3, 25 independent test cases were established to evaluate the simulation performance and image parsing capabilities of *SwarmFoam*. The dataset comprises 10 natural language input cases and 15 multimodal input cases, covering various physical problems in CFD, including single-phase flow, two-phase flow, magnetohydrodynamics, and combustion chemical reactions et al.

**Table 3**
Test Dataset

| Input modality | Problem type | Solver name | Num |
| --- | --- | --- | --- |
| Natural Language | Incompressible | simpleFoam | 1 |
| | | pisoFoam | 2 |
| | | nonNewtonianIcoFoam | 1 |
| | combustion | reactingfoam | 2 |
| | basic | laplacianFoam | 1 |
| | financial | financialFoam | 1 |
| | electromagnetics | mhdFoam | 1 |
| | DNS | dnsFoam | 1 |
| Multimodal | Incompressible | pisoFoam | 2 |
| | | simpleFoam | 3 |
| | | interFoam | 1 |
| | | solidDisplacementFoam | 1 |
| | | icoFoam | 2 |
| | | rhoPimpleFoam | 1 |
| | multiphase | interFoam | 2 |
| | | multiphaseEulerFoam | 1 |
| | compressible | buoyantPimpleFoam | 1 |
| | basic | laplacianFoam | 1 |

### 3.1.2 Evaluation Metrics

The performance of the multi-agent system is evaluated using four metrics: Iterations, Token Usage, Pass Rate, and Cost. Specifically, Iterations is the average number of error corrections required by the multi-agent system to successfully complete one simulation case. It is used to evaluate the error correction capability of the intelligent correction system. Token Usage is the average token expenditure, which includes three parts: input tokens, thinking tokens, and output tokens. Pass Rate represents the proportion of successful completions out of the total number of agent simulation tests, reflecting the system's ability to finish simulation tasks. Cost is the average expenditure for completing one case; we calculate it using the formula from the literature (Gebreab et al., 2026). It is proportional to Token Usage and is related to the pricing model of the LLMs, reflecting the economic efficiency of the system. The specific calculation methods for the four metrics are as follows:

$$\text{Iterations} = \frac{\sum_{i=1}^{n} k_i + m \cdot k_{max}}{m + n} \quad (1)$$

$$\text{Token usage} = \frac{\sum_{i=1}^{m+n} (T_{in,i} + T_{think,i} + T_{out,i})}{m + n} \quad (2)$$

$$\text{Pass rate} = \frac{n}{m + n} \quad (3)$$

$$\text{Cost} = \frac{\sum_{i=1}^{m+n} (T_{in,i} \cdot P_{in} + T_{think,i} \cdot P_{think} + T_{out,i} \cdot P_{out})}{10000(m + n)} \quad (4)$$

Where, $k_i$ is the number of iterations for the i-th successfully run case; $m$ and $n$ are the number of failed and successful tests, respectively; $k_{max}$ is the maximum allowed number of iterations, set to 20 in the experiment; $T_{in,i}$ is the input Token cost for the i-th case; $T_{think,i}$ is the thinking Token cost for the i-th case; $T_{out,i}$ is the output Token cost for the i-th case; $P_{in}$ is the unit price of the LLMs's input Tokens; $P_{think}$ is the unit price of the LLMs's thinking Tokens; and $P_{out}$ is the unit price of the LLMs's output Tokens.

*3.1.3 Model Setup*

In the unified tests, *SwarmFoam* utilized Gemini-2.5-flash for text generation and visual understanding tasks. For the construction of the local knowledge base, HuggingfaceEmbeddings was selected as the embeddings model. The OpenFOAM version used by *SwarmFoam* was OpenFOAM-9. Due to the strict formatting specifications required by OpenFOAM, the temperature parameter of the LLMs was set to 0.01 (a lower value corresponds to lower randomness).

In the comparative tests for the natural language input cases, both *SwarmFoam* and *Foam-agent* (Baseline) utilized DeepSeek-Chat as the LLMs, as *Foam-agent* does not support the Gemini series models.

*3.2 Result and Discussion*

*3.2.1 Overall Performance*

*SwarmFoam* was evaluated on the 25 simulation cases in the dataset, with the results presented in Figure 8. The overall pass rate for the cases was 84% (using Gemini-2.5-flash). Among these, the pass rate for natural language input cases was 80% (Gemini-2.5-flash), while the pass rate for multimodal input cases was 86.67% (Gemini-2.5-flash). Due to the additional image parsing required to process multimodal input, the token consumption and number of iterations for multimodal cases were significantly higher than those for natural language cases. These results indicate that *SwarmFoam* exhibits excellent adaptability to simulation inputs of different modalities.

As *Foam-agent* only accepts natural language input, the 10 natural language cases from the dataset were used for a comparative test (using DeepSeek-Chat for both systems). Although *SwarmFoam* showed no significant difference in pass rate compared to Foam-agent, its average token consumption was substantially lower (-83.85%). This is attributed to the "first-error-priority" correction mechanism introduced in Section 2.3.5. In each iteration, *SwarmFoam* identifies and corrects the single file most likely to have caused the failure. While this may increase the average number of iterations, it effectively reduces the quantity of files rewritten during the correction phase, particularly when handling complex tasks.

A comparison of agent simulation performance on natural language cases using different LLMs revealed that Gemini-2.5-flash (which includes a reasoning mechanism) significantly outperformed DeepSeek-Chat (which lacks a reasoning mechanism) in terms of average iterations (-50.79%) and simulation pass rate (+30%). Due to this added reasoning process, the token usage of the former was higher than that of the latter (+20.86%). Overall, these findings suggest that LLMs equipped with a reasoning mechanism are better suited for handling CFD agent simulation tasks.

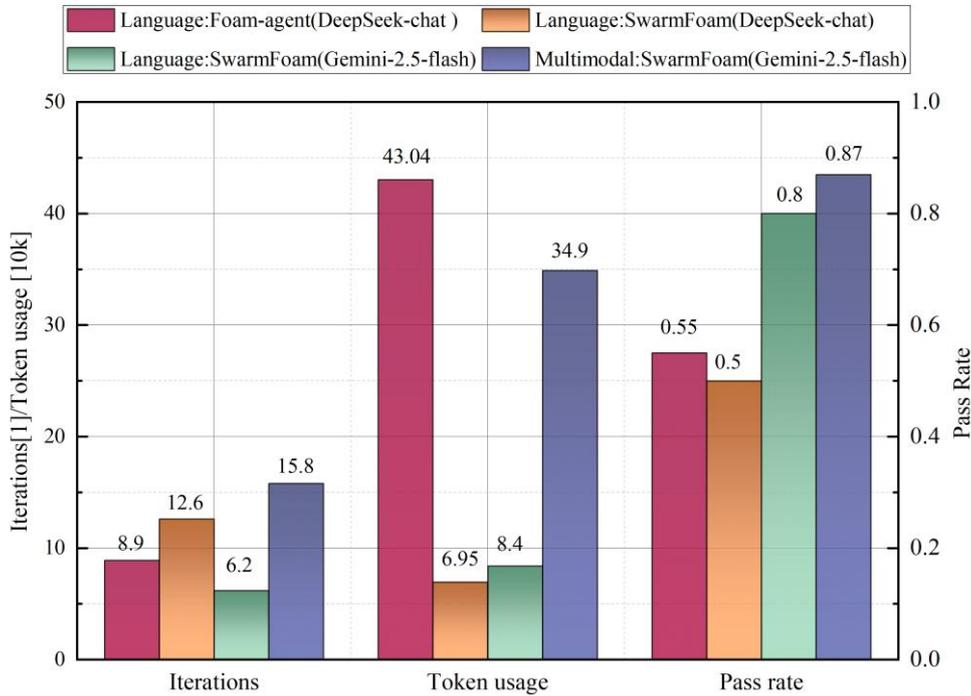

**Fig. 8.** Comparison of overall simulation performance.

*3.2.2 Case Validation*

To validate the accuracy of the *SwarmFoam* agent simulations, three typical CFD cases were selected for discussion: Cavity, Dambreak, and T-junction flow distribution. The natural language inputs for these three cases are shown in Table 4. Specially, the cavity case corresponds to two different geometries and solvers (icoFoam and pisoFoam), in order to demonstrate the adaptability of *SwarmFoam*.

Figure 9 and Figure 10 illustrate the image input and the flow field distribution at $t$=0.5s for the cavity and cavityClipped. It can be observed that for the full cavity, due to the motion of the upper lid, a distinct circulatory flow appears throughout the entire cavity. For the CavityClipped case, due to the missing portion in the bottom-left corner, the circulatory flow is primarily concentrated in the upper part of the cavity. Figure 11 displays the image input for the DamBreak case and the water phase fraction distribution at $t = 9$s. The simulation results show that after the water column collapses, it impacts the bottom obstacle and the tank wall, forming a complex flow structure. Figure 12 presents the image input for the T-junction flow distribution case and the velocity field distribution at $t$ = 100s. After a certain period, the flow velocity inside the pipe reaches a steady state. The main flow is concentrated in the initial velocity direction, while a smaller portion of the fluid exits from the right-hand outlet.

The test results from these three cases demonstrate that the simulation results of *SwarmFoam* show good agreement with the actual results. *SwarmFoam* accurately understood the user requirements and case images, correctly constructed the geometry and mesh, and successfully reproduced the various physical phenomena.

**Table 4**
Natural Language Input

| Case name | Simulation input |
| --- | --- |
| Cavity | do a 2D RANS simulation of incompressible cavity flow using icoFoam |
| CavityClipped | The pisoFoam software was used to conduct a two-dimensional RANS simulation of the incompressible 3/4 cubic cavity flow. The RANS model employed was Kepsilon with a velocity of 1 meters per second. |
| Dambreak | Solve a simplified two-dimensional dam-break problem using interFoam. At the bottom of the water tank, there is a small obstacle. At t = 0s, the water column is allowed to flow freely, and then a water column collapse will occur. |
| T-junction | Use the simpleFoam solver to calculate the 2D flow distribution problem within a T-shaped pipe. The flowing medium inside the pipe is air, with a density of 1 kg/m³ and a dynamic viscosity of 0.003333 kg/m-s. |

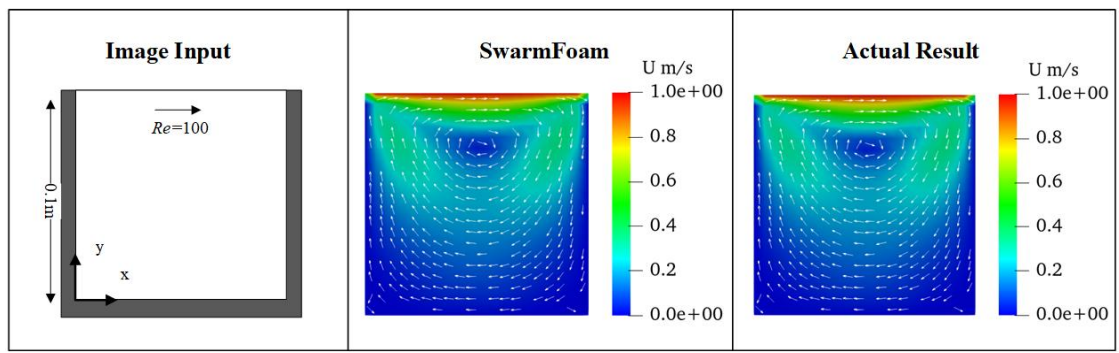

**Fig. 9.** Cavity case (icoFoam)

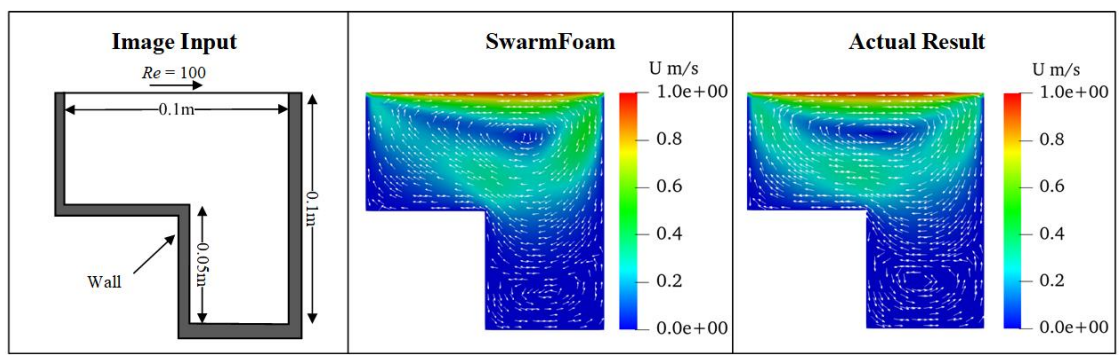

**Fig. 10.** CavityClipped case (pisoFoam)

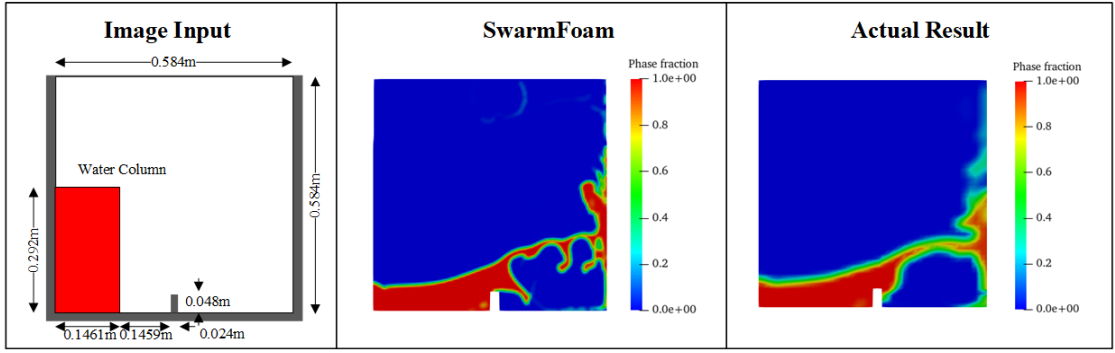

**Fig. 11.** Dambreak case (interFoam)

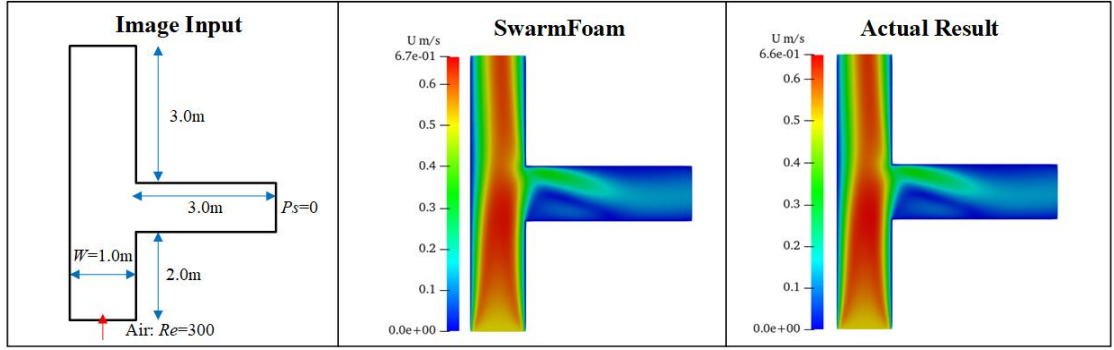

**Fig. 12.** T-junction case (simpleFoam)

*3.2.3 Ablation Study*

The purpose of ablation studies (Gong et al., 2024) is to evaluate the contribution of specific key designs within *SwarmFoam* to the overall simulation performance by removing or replacing them. In this subsection, we conduct ablation studies on the multi-modal perception mechanism and the intelligent error correction mechanism.

1）**Without Multi-modal Perception Mechanism**

The ablation study results for the multi-modal perception mechanism are shown in Figure 13. Method 1 is the approach where image parsing and mesh file generation are performed separately (i.e., decoupled) (using ObservePicture). Method 2 is the approach where image parsing and mesh file generation are performed simultaneously (without ObservePicture).

Since Method 2 does not pre-parse the image, its prompt is shorter. The Observation Agent only executes the DivideTask action and bypasses the ObserverPicture action, resulting in fewer LLMs calls during this stage compared to Method 1. Theoretically, Method 2 should be more economical. However, the actual test results indicate that Method 2 loses a significant amount of geometric and physical information during mesh construction. This leads to a substantially higher average number of iterations (+150.79%) compared to Method 1, with most tasks being forcibly terminated upon reaching the maximum correction limit (20 iterations). As shown in Figure 13, the lower pass rate of Method 2 (-54.02%) paradoxically resulted in nearly double the token usage and average cost. These findings demonstrate that the ObserverPicture action plays a crucial role in enhancing the agent-based simulation performance of *SwarmFoam* for multi-modal cases.

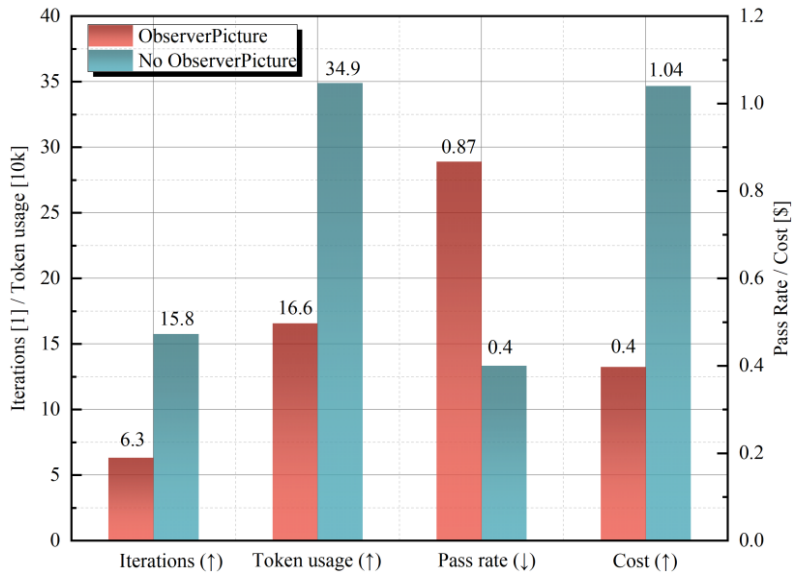

**Fig. 13.** Comparison of simulation performance using different multimodal information parsing methods.

2）**Without Error Correction Mechanism**

The impact of the intelligent error correction mechanism on simulation performance is illustrated in Figure 14. When the Reviewer agent was disabled, the system did not undergo multiple iterations, resulting in lower token usage (-30.95%) and reduced cost for LLMs calls (-40.63%). However, this came at the cost of a 60% reduction in the pass rate. This indicates that, limited by model performance and knowledge scope, the LLMs struggles to complete the agent simulation of complex tasks on the initial attempt and thus requires a reliable error correction mechanism.

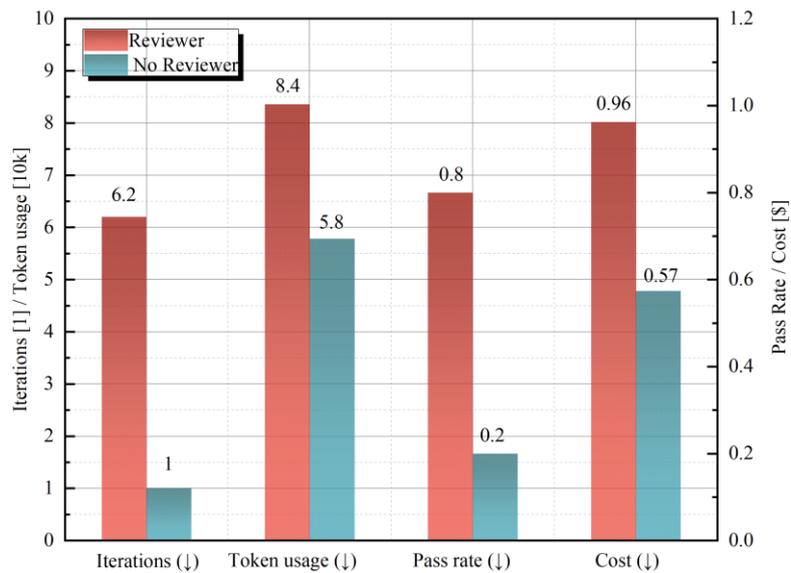

**Fig. 14.** Impact of the error correction mechanism on simulation performance.

*3.2.4 Error Analysis*

As discussed in Section 3.2.3, similar to human experts, LLMs must undergo a series of iterative error correction processes to complete simulation tasks. To facilitate targeted future optimization and upgrades of the agent system, and to simultaneously evaluate the limitations of current general-purpose LLMs in handling CFD tasks, a statistical analysis was conducted on the simulation errors captured by *SwarmFoam* during the case iteration process. The results are shown in Figure 15. For ease of discussion, the simulation errors were classified into five categories:

1）**Configuration Error:** Incorrect configuration of files related to solvers, initial conditions, boundary conditions, and physical properties. As shown in the figure, configuration errors are the predominant form of error for both multimodal input cases (71.1%) and natural language input cases (44.9%). This category of error primarily stems from the limitations of the large language model's (LLM) capabilities and insufficient information from dependent files, such as missing critical parameter settings for the turbulence model or dimensional mismatches in physical quantities.

2）**Geometric Error:** Geometric Error: Errors in 3D geometry or mesh generation, resulting in a failure to pass the mesh quality check. Due to the more complex geometries involved in multimodal input cases, the proportion of geometric errors was 5.1% higher. The mesh generation tool of OpenFOAM relies entirely on the combination of vertex coordinates to define the geometric conditions and perform the meshing. When relying solely on natural language descriptions, it is difficult for the large language model to generate a reliable "blockMeshDict" file.

3）**Missing File:** The absence of files essential for completing the simulation. This error is primarily attributed to two causes. The most significant cause is an improper case file structure planned by the Architect Agent. The second cause is the failure to effectively clean residual artifacts from the output of LLMs, which leads to the OpenFOAM program interpreting certain files as empty.

4）**Grammar Error:** The syntax of the configuration files does not comply with OpenFOAM specifications. This is mainly due to the omission of some OpenFOAM structural symbols in the large language model's output.

5）**Unknown Type:** Insufficient information is available to the Inspector Agent for error correction, including the inability to determine the cause or path of the error.

By integrating mechanisms such as retrieval augmentation, intelligent error correction, and file cleaning, *SwarmFoam* successfully addressed the majority of the aforementioned simulation errors.

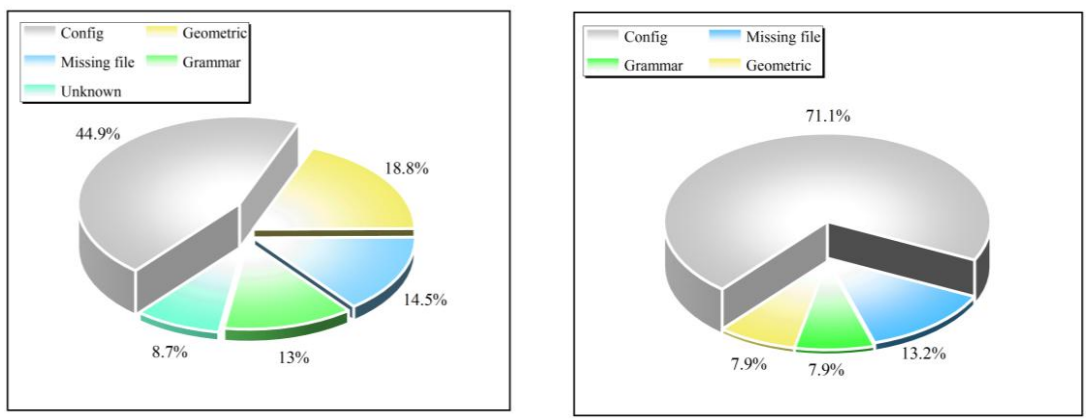

（a）Multimodal Input Case　　　　（b）Natural Langauge Input Case

**Fig. 15.** Statistical analysis of simulation errors.

## 4. Conclusion

CFD is an important means for researchers to understand physical phenomena and evaluate device performance, involving complex processes of geometry construction, mesh generation, and model configuration. Due to these characteristics, CFD tasks have long relied on human experts with specialized training and extensive knowledge in both fluid mechanics and computer science. In this work, the simulation is performed entirely through the cooperation of multiple intelligent agents. Simulation requirements are input in the form of images or high-level instructions and parsed by a multi-modal perception mechanism. Based on the understanding of the simulation task, the quality of the simulation files generated by the agents is enhanced by a RAG system that retrieves local expert knowledge. To evaluate the comprehensive performance of the designed multi-agent system, the proxy system was tested against a simulation dataset spanning multi-modal inputs and covering various physical problems, including single-phase flow, multi-phase flow, and magnetohydrodynamics. The main conclusions drawn in this paper are as follows:

1) The multi-agent framework for CFD based on multi-type LLMs (*SwarmFoam*) can effectively capture case image information and accurately understand users' simulation requirements, successfully realizing intelligent CFD proxy. By combining RAG, multi-modal perception, and an intelligent error correction algorithm, *SwarmFoam* achieved an overall pass rate of 84% across 25 independent test cases.
2) Pre-parsing images to capture geometry and physical information before generating configuration files is significantly important for improving simulation performance. After executing the pre-parsing action (ObserverPicture), the pass rate for multi-modal cases increased by 54.02%, and the cost was reduced by 61.54%.
3) Configuration errors are the main cause of increased iteration counts and higher costs. They accounted for 71.1% and 44.9% of errors in the multi-modal and natural language case tests, respectively. This highlights the necessity for CFD multi-agent proxy systems to possess robust dependent file generation mechanisms and RAG mechanisms.

## 5. Future works

The work of *SwarmFoam* demonstrates the immense potential of multi-agent systems and multi-modal LLMs in realizing intelligent simulation, and its methodology can be extended to other scientific computing fields such as numerical weather prediction, computational structural mechanics, and computational biomechanics. To further inspire the development of CFD agent intelligence and for the future improvement of system, we propose the following recommendations:

- Refining the mesh generation mechanism. General-purpose LLMs still exhibit unreliability when constructing complex geometries and generating meshes, struggling particularly with curved boundaries and unstructured grids. This deficiency may stem from the LLMs' insufficient specialized knowledge of CFD principles and OpenFOAM-specific conventions.
- Enhancing multi-modal perception capabilities. The performance of the multi-modal perception system should be enhanced to enable the parsing of actual engineering drawings. This will likely necessitate the design of more sophisticated multi-modal parsing strategies and the training of specialized visual comprehension models.
- Improving the agent's assessment of simulation result quality. This entails exploring a set of simulation validation criteria to evaluate the fidelity of the results, rather than merely determining the need for configuration correction based on log errors.

## Acknowledgement

This research was supported by the CNNC Leading Innovation Fund (Project Number: CNNC-LCKY-2023-047), the Sichuan Science and Technology Program (Grant No. 2025ZNSFSC0082), the State Key Laboratory of Advanced Nuclear Energy Technology, Nuclear Power Institute of China (Grant No.YNSW-0224-0202-04-01), the "Young Talents Program" of China National Nuclear Corporation (CNNC) under the project entitled "Study on Bubble Transport Mechanisms and Two-Phase Transient Phenomena in the Fuel Solution of the Medical Isotope Test Reactor".

## Appendix

### A.1 Prompt Template

All agents utilize a standardized prompt template. Each agent modifies this template to some extent based on the specific requirements of its task. The partial prompts of *SwarmFoam* are shown in the figures below.

---

**ObserverPicture Prompt**

**\<Your task\>**
Based on the description of the simulation object and the pictures, analyze the geometric and physical information.

**\<task description\>**
{cfd_example_describe}

**\<Reference information\>**
The relevant content for completing the current task

**\<Output requirement\>**
Return ```The information of the example picture is as follows:
**Geometric description:** This section contains three pieces of information. The first part is the geometric description. The second part is the vertex coordinates. The third part contains the coordinate information of edges or faces.
**Physical description:** This part includes the physical information in the analyzed picture, such as flow velocity, flow direction, temperature, pressure, etc.```
if no picture, both return "None"

Keep "Geometric description:" and "Physical description:" in your return.
Do not output any other content or identifier.
Do not include special symbols in the reply.
Do not have any other identifiers in the begin and end of the output file.

## DivideTasks Prompt

**<Your task>**
According to the cfd_example_describe in the task description, the user requirements need to be decomposed into simulation tasks and post-processing tasks.

**<task description>**
{cfd_example_describe}

**<Reference information>**
The relevant content for completing the current task

**<Output requirement>**
Return The tasks is as follows:```
**simulation tasks:** The actual simulation requirements you have obtained.
**post-processing tasks:** The post-processing requirements in the description. If the description does not require the post-processing task, content after "post-processing tasks:" to be "None".```

Do not output any other content or identifier.
Do not include special symbols in the reply.
Do not have any other identifiers in the begin and end of the output file.

## HandleError Prompt

**<Your task>**
Based on the error message from OpenFOAM, determine the type of error (note) error type must be one of the follows:format error, Missing file. The "Missing file" function only considers the files within the "0", "system" and "constant" folders, as well as the "Allrun" file.

**<error message>**
{errors}

**<Existing files>**
{file_list}

**<Error type explain>**
**Format error:** Some configuration files do not comply with the OpenFOAM standards (such as having incorrect formats) or are improperly configured, or missing file keyword.
**Missing error:** A certain file is missing in the example directory, which don't in the existing files.

**<Output requirement>**
Return error type. Do not output any other content.

## WriteFoamFile Prompt

**\<Your task\>**

For input requirement in user requirement, If there are similar input files, reference the similar examples (for reference only) to create the input files.

If there is no reference, complete the creation directly based on the description for OpenFOAM-9.

Simulation task information is in task description. The physical information of the simulation task is in Physical information.

The geometric information of the simulation object is in Geometrical information or base the picture.

The dimension of physical quantities are consistent with those in the "similar file".

If the reference example uses the same solver as the current one, the dimensions should be consistent.

The current version of OpenFOAM is 9.

**\<User requirement\>**
{requirement}

**\<task description\>**
{CFD_task}

\<Physical information\>
{physical_information}

\<Geometrical information\>
{geometrical_information}

\<solver help\>
Some dimension or naming suggestions
{solver}

\<similar file\>
for reference only
{similar_file}

**\<dependent file\>**
The "dependent file" refers to the files that are dependencies of the target file.

Ensure that the boundary settings of the current file and the dependent files are consistent, and that the parameter settings are also matched.
{associated_file}

**\<Output requirement\>**
Return ```The input file is:The content you have written```
Do not output any other content or identifier.
Do not include special symbols in the reply.
Do not have any other identifiers in the begin and end of the output file.